\documentclass[12pt]{article}

 \newcommand{\newc}{\newcommand}
\newc{\beq}{\begin{equation}} \newc{\eeq}{\end{equation}}
\newc{\bea}{\begin{array}} \newc{\eea}{\end{array}}
\newc{\ri}{{\mathrm i}}
\newc{\bW}{{\mathbf W}}
\newc{\bR}{{\mathbf R}}
\newc{\bN}{{\mathbf N}}
\newc{\Psibar}{\overline\Psi}
\newc{\w}{{\bf w}}
\newc{\E}{{\mathbf{E}}}
\newc{\bp}{{\bf p}}
\newc{\ta}{\tilde a}
\newc{\bV}{{\bf V}}
\newc{\bfV}{{\bf V}}
\newc{\bfG}{{\bf G}}
\newc{\bx}{{\bf x}}
\newc{\bu}{{\bf u}}
\newc{\bP}{{\bf P}}
\newc{\bJ}{{\bf J}}
\newc{\bK}{{\bf K}}
\newc{\pd}{{\partial}}
\newc{\ti}{{\times}}
\newc{\bA}{{\bf A}}
\newc{\bE}{{\bf E}}
\newc{\bfn}{{\bf\nabla}}
\newc{\ho}{\hookrightarrow}
\newc{\ra}{\rightarrow}
\newc{\bv}{{\bf v}}
\newc{\bb}{{\bf b}}
\newc{\bc}{{\bf c}}
\newc{\bd}{{\bf d}}
\newc{\tbb}{\tilde{\bf b}}
\newc{\tbc}{\tilde{\bf c}}
\newc{\tbd}{\tilde{\bf d}}
\newc{\bz}{{\bf 0}}
\newc{\bun}{{\bf 1}}
\newc{\bL}{{\bf L}}
\newc{\bS}{{\bf S}}
\newc{\bB}{{\bf B}}
\newc{\br}{{\bf r}}
\newc{\sig}{{\mathbf\sigma}}
\newc{\eg}{{\it e.g.\ }}
\newc{\bpi}{{\mathbf\pi}}
\newc{\ie}{{\it i.e.\ }}
\newc{\etal}{{\it et al}}

\def\JPA#1#2#3#4{#2 #1 {\em J. Phys. A: Math. Gen.} {\bf #3} #4}
\def\FP#1#2#3#4{#2 #1 {\em Fortschr. Phys.} {\bf #3} #4}

\def\AM#1#2#3#4{#2 #1 {\em Ann. Math.} {\bf #3} #4}
 \def\APP#1#2#3#4{#2 #1 {\em Acta Phys. Pol.} {\bf #3} #4}

\def\NCB#1#2#3#4{#2 #1 {\em Nuov. Cim.} {\bf #3 B} #4}
\def\NCA#1#2#3#4{#2 #1 {\em Nuov. Cim.} {\bf #3 A} #4}

\def\IJTP#1#2#3#4{#2 #1 {\em Int. J. Theor. Phys.} {\bf #3} #4}
\def\CMP#1#2#3#4{#2 #1 {\em Comm. Math. Phys.} {\bf #3} #4}

\def\JMP#1#2#3#4{#2 #1 {\em J. Math. Phys.} {\bf #3} #4}
\def\PNASUS#1#2#3#4{#2 #1 {\em Proc. Nat. Acad. Sci. U.S.} {\bf #3} #4}
\def\TMP#1#2#3#4{#2 #1 {\em Theor. Math. Phys.} {\bf #3} #4}

\def\TMP#1#2#3#4{#2 #1 {\em Theor. Math. Phys.} {\bf #3} #4}

\def\DUK#1#2#3#4{#2 #1 {\em Duke Math. J.} {\bf #3} #4}
\def\REW#1#2#3#4{#2 #1 {\em Rev. Math. Phys} {\bf #3} #4}
 \catcode`@=11 \long
\def\@caption#1[#2]#3{\par\addcontentsline{\csname
ext@#1\endcsname}{#1} {\protect\numberline{\csname
the#1\endcsname}{\ignorespaces #2}} \begingroup \small
\@parboxrestore \@makecaption{\csname fnum@#1\endcsname}
{\ignorespaces #3}\par \endgroup} \catcode`@=12

\begin{document} \begin{titlepage} \vskip 2cm
\begin{center} {\Large\bf Galilei invariant theories.\\I. Constructions of indecomposable
 finite-dimensional representations of the homogeneous Galilei group: directly and via contractions}
\footnote{E-mail: {\tt montigny@phys.ualberta.ca},\ \
{\tt niederle@fzu.cz},\ \ {\tt nikitin@imath.kiev.ua} }
\vskip 3cm {\bf M. de Montigny$^{a,b}$, J. Niederle$^c$, \\ and
A.G. Nikitin$^d$ }
\vskip 5pt
{\sl $^a$Campus Saint-Jean, University of Alberta, 8406 - 91
Street, Edmonton, Alberta, Canada T6C 4G9\\}
\vskip 2pt
{\sl $^b$Theoretical Physics Institute, University of Alberta,
Edmonton, Alberta, \\Canada T6G 2J1 \\}
\vskip 2pt
 {\sl $^c$Institute of Physics of the
Academy of Sciences of the Czech Republic, Na Slovance 2, 18221
Prague 8, Czech Republic}
\vskip 2pt
 {\sl $^d$Institute of
Mathematics, National Academy of Sciences of Ukraine,\\ 3
Tereshchenkivs'ka Street, Kyiv-4, Ukraine, 01601\\}
\end{center}
 \vskip .5cm \rm

\begin{abstract} All indecomposable finite-dimensional representations of the
homogeneous Galilei group  which when restricted to the rotation
subgroup are decomposed to spin 0, 1/2 and 1
representations are constructed and classified. These representations are also obtained via contractions 
of the corresponding representations of the Lorentz group. 
Finally the obtained representations are used to derive a general Pauli 
anomalous interaction term and Darwin and spin-orbit couplings of a Galilean particle 
interacting with an external electric field.
 \end{abstract}

 \end{titlepage}

\setcounter{footnote}{0} \setcounter{page}{1}
\setcounter{section}{0}

\section{Introduction}

Unlike what most physicists might think of, the mathematical
structure of the representations of the Galilei group is in many
respects more complex and sophisticated  than that of their
relativistic counterparts.
This is perhaps also the reason why unitary irreducible representations of the
Poincar\'e group -- the symmetry group of
special theory of relativity -- has been known nearly  twenty years earlier
than those of the Galilei group,
even though the Galilei principle of relativity
 was discovered several centuries prior to the Einstein one.

The Galilei group and its representations are described in the L\'evy-Leblond masterful survey
\cite{levyleblond} written
  almost thirty five years ago. They form the
group-theoretical basis for description of various physical
predictions in non-relativistic classical mechanics and
electrodynamics and in non-relativistic
quantum mechanics as well (see also a more recent review \cite{Cassinelli}) . 
These predictions are generally more
than just some approximations to the relativistic results
since only comparison of the predictions based on the Galilei group
and its representations with those on the Poincar\'e group can indicate which predictions have the 
origin in the non-relativistic
and which in the relativistic one. For instance, it was shown in \cite{Wigner}, \cite{bargman}
 and by L\'evy-Leblond in \cite{levyleblond},
  \cite{ll1967}, \cite{lebellac}
that the concept of spin of particles and of magnetic moments of
particles have the origin already in the Galilean
non-relativistic quantum mechanics and not as stated in many
textbooks as a consequence of relativistic effects.

In the Galilei invariant framework we shall consider first free particles and then particles interacting 
with external fields. The free particles can be described either by  representations of the Galilei group 
(irreducible in the case of elementary particles and indecomposable for particles with internal structures)
 or equivalently, by Galilean invariant wave equations. For interacting particles the wave equations are 
 more appropriate since they allow to introduce interactions.

There exist three approaches how to formulate the Galilei invariant
wave equations. The first is based on the fact that the Galilei group and the
Galilei invariant equations can be obtained from the Poincar\'e
group and Poincar\'e invariant equations
respectively by a limiting procedure -- the so-called
Inon\"u-Wigner contraction \cite{contraction}.

The second approach consists of writing first the Lorentz-invariant equations in
the $(4+1)$-dimensional space-time, and then
projecting them down to the $(3+1)$-dimensional Newtonian space-time
using the fact that the extended Galilei algebra in $(3+1)$
dimensions is a subalgebra of the Poincar\'e algebra in $(4+1)$
dimensions, for connection of representations of these algebras see \cite{fushchichnikitin5d}. 
This approach, referred here
as a projecting one, has been developed
in \cite{takahashi}-\cite{kapuscik} (see also \cite{nikitin}).

The third way to construct Galilei invariant theories consists in
searching for this theories directly
using the requirement of Galilei invariance  and knowledge of representations of the Galilei group.
We shall show in this paper that, in many respects,
this latest approach is the most powerful
and comprehensive.
 Moreover, it allows us to construct such consistent
Galilei-invariant equations of motion which are very difficult
 to derive using the contraction or projection methods.
 On the other hand, the direct search for wave equations invariant with respect to the inhomogeneous 
 Galilei group
 presupposes a good knowledge of the indecomposable finite-dimensional  representations
 of its homogeneous Galilei subgroup, which has not been available till now.

Our article serves the following four aims: (1) to describe all indecomposable
finite-dimensional
representations of the homogeneous Galilei group that are defined on spinor, scalar
and vector representation spaces; (2) to specify all Galilei invariant
 bilinear forms, which facilitate derivation of various
non-linear Galilei wave equations; (3) to specify the reducible representations of the Lorentz group 
which lead to the found representations of the
Galilei group via contractions;
(4) to determine the Galilean
spinorial wave equations which include the Pauli anomalous terms.

In the next section we define indecomposable finite-dimensional representations of the homogeneous 
Galilei group in general.

In Sections 3 and 4 some of these representations are constructed explicitly, namely all those
 which when restricted to representations
of the rotation subgroup of the homogeneous Galilei group are decomposed to spin 0, 1/2 and 1
representations. In addition we present here also the complete list of bilinear forms
invariant with respect to all found representations of the Galilei group. Section 5 contains various 
examples of the Galilean vectors.
In Section 6 we obtain the previously found representations of the homogeneous Galilei group via the 
In\"on\"u-Wigner contraction of the corresponding representations of the Lorentz group.
In Section 7 the found representations are used to deduce the most general form
of the Pauli interaction term which can be added to the
Galilei invariant equation for spinors. Subsection 7.3 presents
a simple Galilean system which describes the Darwin and spin-orbital interactions of particles with an 
external electric field.
Finally, Section 8 is devoted to discussions of the obtained results.

\section{Definitions and properties of the Galilei group and its Lie algebra}

The Galilei group $G(1,3)$ consists of the following transformations of time variable $t$ and of 
space variables ${\bf x} =
(x_1,x_2,x_3) $:
\beq\label{11}\bea{l} t\to t'=t+a,\\{\bf x}\to {\bf x}'={\bf R}{\bf x} +{\bf v}t+\bf b\eea\eeq
where $a,{\bf b}$ and ${\bf v}$ are real parameters of time translation, space translations and pure 
Galilei transformations respectively, and matrix $\bf R$ specifies rotations determined by the three 
parameters ${\theta}_1, \theta_2, \theta_3$.

The Galilei group includes a subgroup leaving invariant a point $\textbf{x}=(0,0,0)$ at time $t=0$. 
It is formed by all space rotations and pure Galilei transformations, i.e., by transformations (\ref{11}) 
with $a={\bf b}\equiv 0$. This subgroup is said to be {\it the homogeneous Galilei group} $HG(1,3)$. 
It is a semi-direct product of the three-parameter commutative group of pure Galilei transformations 
with the rotation group.  Thus this group is not compact and does not have unitary finite-dimensional 
representations.

The Galilei group $G(1,3)$  is a semidirect product of its  Abelian subgroup generated by time and space 
translations with the homogeneous Galilei group $HG(1,3)$ .

Unitary representations of Galilei group which are ordinary ones were described
by In\"on\"u and Wigner \cite{Wigner} in 1952, whereas those which are ray by  Bargmann \cite{bargman} 
in 1954. A nice
review of these representations can be found in \cite{levyleblond}, see also \cite{fuschich}.

However, a decisive role in the description of various finite-component
Galilei invariant  equations is played by finite-dimensional
representations of the homogeneous Galilei group $HG(1,3)$.
They where studied first apparently
 in paper \cite{nahas} but have not been classified till now.

 Let us remind that the representations of $HG(1,3)$
 induce ray representations of the Galilei group $G(1,3)$ as well as ordinary representations of the 
 extended Galilei group $G_m(1,3)$ which is a central extension of the Galilei group via a one-parameter 
 subgroup. Both of them are realized in the space of (square integrable) $n$-component functions 
 $\Psi(t,{\bf x})$ which for any transformation (\ref{11})
cotransform  in the following way \cite{levyleblond}
\beq\label{cova}
\Psi({\bf x}, t)\to\Psi'({\bf x'}, t')=e^{\ri f({\bf x},
t)}T\Psi({\bf x}, t)\eeq
where $\Psi({\bf x}, t)=\texttt{column}(\Psi_1({\bf x}, t),\Psi_2({\bf x},
t),
\cdots,\Psi_n({\bf x}, t))$ are $n$-component vectors from the representation
space, $T$ are
$n\times n$ numerical matrices depending on transformation parameters ${\bf v}$ and 
${\mbox{\boldmath$\theta$\unboldmath}}=(\theta_1, \theta_2, \theta_3)$,
\[f({\bf x}, t)=m\left({\bf v}
\cdot{\bf x}+\frac{v^2}{2}t+c\right)\]
is a phase which includes two parameters: $m$ and $c$. For $m=0$ the central extension is trivial and 
representations (\ref{cova}) are ordinary representations of $G(1,3)$. For $m$ different from zero the 
central extension is nontrivial, and transformations (\ref{cova}) realize ray representations of the 
Galilei group $G(1,3)$ and ordinary representations of $G_m(1,3)$. Moreover, the
transformation matrices $T$ realize a finite-dimensional representations
of the homogeneous Galilei group $HG(1,3)$.

Let us mention that realizations (\ref{cova}) are precisely those which are used in quantum mechanics
and quantum field theory. These realizations are also essential to
construct wave equations invariant w.r.t. the Galilei group \cite{nikitin}.

Taking expressions (\ref{cova}) corresponding to the infinitesimal transformations (\ref{11}) and 
treating $c$ as an additional transformation parameter we can calculate the 11-dimensional Lie algebra 
of the extended Galilei group.
Basis elements of this algebra are of the following form
 \beq\label{cov}\bea{l} P_0=\ri\partial_0,\ \  P_a=-\ri\partial_a,\ M=Im\\
 J_a=-\ri\varepsilon_{abc}x_b\partial_c+S_a,\\
 K_a=-\ri x_0\partial_a-mx_a+\eta_a,\end{array}\eeq
  where indices $a, b$ and $c$ run over the values $1,2,3$, $I$ is the unit matrix, and $S_a$ and 
  $\eta_a$ are matrices which
satisfy the following commutation relations:
\beq\label{e3}
[S_a,S_b]=\ri\varepsilon_{abc}S_c,\eeq
 \beq\label{e32}   {[}\eta_a,S_b{]}=\ri\varepsilon_{abc}\eta_c,\eeq
\beq\label{e31}{[}\eta_a,\eta_b{]}=0,\eeq
that is, they form a basis of the Lie algebra $\textsl {h}\textsl{g}(1,3)$ of the
homogeneous Galilei group $HG(1,3)$. This algebra is isomorphic to the Lie algebra
 $hg(1,3)$ of the Euclidean group.

Conditions (\ref{e3}) -- (\ref{e31}) are necessary and sufficient in order generators
(\ref{cov}) form a basis of a Lie algebra namely the extended Galilei algebra satisfying the following 
relations:
 \beq
\label{galal}\bea{ll} [J_a,
J_b]=\ri\epsilon_{abc}J_c,\quad & {[J_a, K_b]}=\ri\epsilon_{abc}K_c,
\\
{[J_a, P_b]}=\ri\epsilon_{abc}P_c,\quad & {[K_a, P_0]}=\ri P_a,\\
{[K_a,P_b]}=\ri\delta_{ab} M, \quad &{[K_a,K_b]}=0,\\{[P_a,P_b]}=0,\quad &{[P_0,P_a]}=0,\\
{[M,P_0]}={[M,P_a]}=&{[M,J_a]}={[M,K_a]}=0.  \eea\eeq

Unfortunately, the problem of a
complete classification of non-equivalent finite-dimensional
realizations of algebra (\ref{e3})-(\ref{e31}) appears to be an unsolvable
`wild' algebraic problem.  However we shall show that this
problem can be completely solved in the two important particular
cases: for the purely spinor representations and vector representations.

\section{Spinor representations}

The Lie algebra $hg(1,3)$, defined by relations (\ref{e3})-(\ref{e31}),
includes the
subalgebra $so(3)$ spanned on the basis elements $S_1, S_2$ and
$S_3$. Without loss of generality we suppose that representations of this
subalgebra are Hermitian and completely reducible and shall search for
representations of $hg(1,3)$ in the $so(3)$ basis in which the Casimir
operator of $so(3)$ is diagonal.

Irreducible representations of $so(3)$
are labelled by integers or half-integers $s$.
Let $\tilde s$ be the highest value of $s$ which appears in the decomposition of a reducible 
representation of $so(3) $ subduced by the indecomposable representation of $hg(1,3)$.
We shall call the related  carrier space of this representation of $hg(1,3)$ {\it the representation 
space of spin
$\tilde s$}.

Consider finite-dimensional indecomposable representations of algebra $hg(1,3)$ of
spin $\tilde s=\frac12$. Then the corresponding matrices $S_1, S_2$ and $S_3$  can be decomposed to a 
direct
sum of the irreducible representations $D(1/2)$ of algebra $so(3)$:
\beq\label{o3} S_a=\frac12I_{n\times n}\otimes \sigma_a,\eeq
where
$I_{n\times n}$ is the $n\times n$ unit matrix with a finite $n$
and $\sigma_a$ are the usual Pauli matrices:
\[
\sigma_1=\left (
\bea{cc}
 0 & 1 \\
 1 & 0\eea\right ),\
\sigma_2=\left (
\bea{cc}
 0 & -\ri \\
 \ri & 0\eea\right ),\
\sigma_3=\left (
\bea{cc}
 1 & 0 \\
 0 & -1\eea\right ).
\]

From (\ref{o3}) and
(\ref{e32}), the generic form of the related matrices $\eta_a$ is:
\beq\label{eta}\eta_a=A_{n\times n}\otimes \sigma_a,\eeq
 where
$A_{n\times n}$ is an $n\times n$ matrix.
The commutativity of matrices (\ref{eta}) leads to the nilpotency condition
\beq\label{nilpotent}A_{n\times n}^2=0.\eeq
Thus, without any loss of generality,
 $A_{n\times n}$ may be expressed as a direct sum of $2\times2$ Jordan
 cells and zero
 matrix.

Because of (\ref{nilpotent}) there exist only two different
indecomposable representations of algebra $hg(1,3)$ defined on spin
1/2 carrier space:
 \[ S_a=\frac12\sigma_a,\qquad \eta_a=0,\]
 and
\beq\label{02}S_a=\frac12\left(\bea{cc}\sigma_a&0\\0&\sigma_a\eea\right),\qquad
\eta_a=\frac{1}2\left(\bea{cc}0&0\\\sigma_a&0\eea\right).\eeq
 The corresponding vectors from the representation space are two-component
 spinors $\varphi({\bf x}, t)$, as well as four-component
 bispinors $\Psi=\left(\bea{l}\varphi_1({\bf x}, t)\\\varphi_2({\bf x}, t)\eea\right)$
 with two-component $\varphi_1$ and $\varphi_2$, respectively.
 When $t$ and $\bf x$ undergo a
Galilean transformation (\ref{11}), then
$\varphi$ cotransforms as
\[
\varphi({\bf x}, t)\to\varphi'({\bf x'}, t')=e^{\ri m\left({\bf
v}\cdot{\bf x}+{\bf v}^2t/2+c\right)}\left(\cos\frac\theta2+i\frac{
{{\mbox{\boldmath$\sigma$\unboldmath}}}\cdot{{\mbox{\boldmath$\theta$\unboldmath}}}}
\theta\sin\frac\theta2\right) \varphi({\bf x}, t)\]
while the transformation law for components of Galilean bispinor is
\[\bea{l}\displaystyle \varphi_1({\bf x}, t)\to\varphi_1'({\bf x'}, t')=e^{\ri m\left({\bf
v}\cdot{\bf x}+{\bf v}^2t/2+c\right)}\left(\cos\frac\theta2+
i\frac{{\mbox{\boldmath$\sigma$\unboldmath}}
\cdot{{\mbox{\boldmath$\theta$\unboldmath}}}}
\theta\sin\frac\theta2\right)\varphi_1({\bf x}, t),\\\\\displaystyle
\varphi_2({\bf x}, t)\to\varphi_2'({\bf x'}, t')=
e^{\ri m\left({\bf
v}\cdot{\bf x}+{\bf v}^2t/2+c\right)}\left(\left(\cos\frac\theta2+
i\frac{{{\mbox{\boldmath$\sigma$\unboldmath}}}\cdot{{\mbox{\boldmath$\theta$\unboldmath}}}}
\theta\sin\frac\theta2\right)\varphi_2({\bf x},
t)\right.\\\\\left.+\left(\frac{\ri}2{{\mbox{\boldmath$\sigma$
\unboldmath}}
\cdot {\bf v}}\cos\frac\theta2-
({{\mbox{\boldmath$\theta$\unboldmath}}}\cdot{\bf v}+
i{\mbox{\boldmath$\sigma$\unboldmath}}\cdot
{\mbox{\boldmath$\theta$\unboldmath}}\times{\bf v})\frac{\sin\frac\theta2}
{\theta}\right)\varphi_1({\bf x}, t)\right).
\eea
\]

We use the notation $\theta=\sqrt{\theta_1^2+\theta_2^2+\theta_3^2}$. 
Invariants of these transformations which are independent on $t$ and ${\bf x}$ are
 arbitrary functions of $\varphi^\dag\varphi$,   $\varphi^\dag_1\varphi_1$ and 
 $\varphi^\dag_1\varphi_2+\varphi^\dag_2\varphi_1$.

\section{Vector representations}

In this section, we examine the finite-dimensional indecomposable representations of the algebra $hg(1,3)$
defined on vector, or spin-one, representation spaces.
The corresponding matrices $S_a$ can be expressed as direct sums of spin-one and spin-zero
 matrices:
\beq\label{s}S_a=\left(\bea{cc}I_{n\times n}\otimes
s_a&\cdot
\\
\cdot&\bz_{m\times m}
\eea\right).\eeq
The symbols $I_{n\times n}$ and $\bz_{m\times m}$ denote the $n\times n$ unit
matrix and $m\times m$ zero matrix respectively, $s_a$ $(a=1,2,3)$ are $3\times 3$
matrices of spin equal to one for which we choose the following realization:
\beq\label{spin} s_1= \left (\bea{ccc}0 &0 &0 \\ 0& 0& -\ri \\
0 & \ri & 0\eea\right ),\quad s_2= \left (\bea{ccc} 0&0 & \ri
\\0 &0 & 0\\ -\ri &0 & 0\eea\right ),\quad s_3= \left (\bea{ccc}0 & -\ri
&0
\\ \ri & 0&0 \\ 0& 0&0 \eea\right )\eeq

The general form of matrices $\eta_a$ which satisfy relations (\ref{e32})
with matrices (\ref{s}) is given by the following formulae (see, e.g., \cite{gelfand}):
\beq\label{eta1}\eta_a
=\left(\bea{cc}A\otimes s_a&B\otimes k^\dag_a\\C\otimes k_a
&\bz_{m\times m}\eea\right)\eeq  $A$, $B$ and $C$ are matrices
of dimension $n\times n$, $n\times m$ and $m\times n$
respectively, $k_a$ are $1\times3$ matrices of the form
\beq\label{k}
k_1=\left (\ri, 0, 0\right),\qquad k_2=\left (0, \ri, 0\right),
\qquad k_3=\left (0, 0, \ri\right).\eeq

The matrices (\ref{s}) and (\ref{eta1}) satisfy conditions (\ref{e32}) with
any $A, B$ and $C$. Substituting (\ref{eta1}) into
(\ref{e31}) and using the relations
\[\bea{l}s_ak^\dag_b=i\varepsilon_{abc} k^\dag_c,\
k_as_b=i\varepsilon_{abc} k_c,\\
{[}s_a,s_b]=k^\dag_ak_b-k^\dag_bk_a=i\varepsilon_{abc}s_c,\\
k_ak^\dag_b-k_bk^\dag_a=0\eea\]
we obtain the following equations for matrices $A,\ B$ and $C$:
\beq\label{A1}A^2+BC=0,\eeq\beq\label{A2}CA=0,\ \ AB=0.\eeq

The solution of the matrix problem defined by equations (\ref{A1}) and  (\ref{A2}) is relatively easily to
handle. Namely, if we multiply equation (\ref{A1}) by $A$ and use equation (\ref{A2}), then
we obtain the condition $A^3=0$  so that $A$ is a nilpotent matrix
whose nilpotency index $N$ satisfies the condition $N\leq 3$. This implies that
$BC$ is a nilpotent
matrix with index of nilpotency equal to 2. Thus $A$ can be
represented as a direct sum of the Jordan cells of dimension 3,\ 2 and
zero matrices. We can prove that in order to obtain
indecomposable representations of algebra (\ref{e3}), (\ref{e31})
it is necessary to restrict ourselves to the case of the
indecomposable matrices $A$, and so there are three possibilities
\beq\label{A3}A=\left(\bea{ccc}0&0&0\\1&0&0\\0&1&0\end{array}\right),
\ A=\left(\bea{cc}0&0\\1&0\end{array}\right),\ A=0.\eeq

We can prove that for indecomposable matrices $\eta$
dimension $m$ and dimension $n$ satisfy the condition $-1\leq (n-m)\leq2, \ n\leq3$.
Matrices $\eta_a$ are defined up to a constant multiplier, which we fix imposing the condition
\beq\label{A33} \texttt{Trace}(\eta^\dag_a\eta_a)\equiv\texttt{Trace}(\eta^\dag_1\eta_1+\eta^\dag_2\eta_2+
\eta^\dag_3\eta_3)=6(n-1)+3(m+\delta_{3n}).\eeq

By going over all admissible values of $n, \ m$ and using relations (\ref{A3})
 we find that up to equivalence
transformations there exist ten solutions of equations (\ref{A2}), which can be labelled by
triplets of numbers $n,m,\lambda$ where $\lambda =\frac13\texttt{Trace}\left( 
(S_aS_a-2)\eta^\dag_a\eta_a\right)$ takes the values
$$\lambda=\left\{\bea{l}0\ \texttt{if}\ m=0,\\
1\ \texttt{if}\ m=2\ \texttt{or}\ n-m=2,\\ 0, 1\ \texttt{if}\ m=1, n\neq3.\eea\right.$$
We shall
not present them here but write directly the corresponding matrices $S_a$
(\ref{s}) and $\eta_a$ (\ref{eta1}),
 which are collected in Table
\ref{vector}.

\begin{table}
\begin{tabular}{|c|c|c|}\hline
Representation& $S_a$ & $\eta_a$ \\
$D(n,m,\lambda)$&&\\
\hline
 & & \\
$D(0,0,0)$ & $0$ & $0$ \\ & & \\
$D(1,0,0)$ & $s_a$ & $\bz_{3\times3}$ \\ & & \\
$D(1,1,0)$ & $\left(\bea{cc}s_a&\bz_{3\times1}\\
\bz_{1\times3}&0\eea\right)$ & $\left(\bea{cc}\bz_{3\times3}&
\bz_{3\times1}\\
k_a&0\eea\right)$ \\ & & \\
$D(1,1,1)$ & $\left(\bea{cc}s_a&\bz_{3\times1}\\
\bz_{1\times3}&0\eea\right)$ & $\left(\bea{cc}\bz_{3\times3}&k^\dag_a\\
\bz_{1\times3}&0\eea\right)$ \\ & & \\
$D(1,2,1)$ & $\left(\bea{ccc}s_a&\bz_{1\times3}&\bz_{1\times3}\\
\bz_{3\times1}&0&0\\
    \bz_{3\times1}&0&0\eea\right)$ &
  $\left(\bea{ccc}\bz_{3\times3}&k^\dag_a&\bz_{3\times1}\\
  \bz_{1\times3}&0&0\\
    k_a&0&0\eea\right)$ \\ & & \\
$D(2,0,0)$ & $\left(\bea{cc}s_a&\bz_{3\times3}\\
\bz_{3\times3}&s_a\eea\right)$ &
 $\left(\bea{cc}\bz_{3\times3}&\bz_{3\times3}
 \\s_a&\bz_{3\times3}\eea\right)$ \\ & & \\
$D(2,1,0)$ & $\left(\bea{ccc}s_a&\bz_{3\times3}&\bz_{3\times1}
\\\bz_{3\times3}&s_a&\bz_{3\times1}\\\bz_{1\times3}&\bz_{1\times3}&0\eea
\right)$ &
 $\left(\bea{ccc}\bz_{3\times3}&\bz_{3\times3}&\bz_{3\times1}\\s_a&
 \bz_{3\times3}
    &\bz_{3\times1}\\k_a&\bz_{1\times3}&0\eea\right)$ \\ & & \\
$D(2,1,1)$ &  $\left(\bea{ccc}s_a&\bz_{3\times3}&\bz_{3\times1}
\\\bz_{3\times3}&s_a&\bz_{3\times1}\\\bz_{1\times3}&\bz_{1\times3}&0\eea
\right)$ &$\left(\bea{ccc}\bz_{3\times3}&s_a&k^\dag_a\\\bz_{3\times3}&
\bz_{3\times3}&\bz_{3\times1}\\\bz_{1\times3}&\bz_{1\times3}&0\eea\right)$
  \\  & & \\
$D(2,2,1)$ & $\left(\bea{cccc}s_a&\bz_{3\times 3}&\bz_{3\times
1}&\bz_{3\times 1}\\\bz_{3\times 3}&s_a&\bz_{3\times
1}&\bz_{3\times 1}\\\bz_{1\times 3}&\bz_{1\times 3}&0&0\\
\bz_{1\times 3}&\bz_{1\times 3}&0&0\eea\right)$ &
 $\left(\bea{cccc}\bz_{3\times 3}&\bz_{3\times
3}&\bz_{3\times 1}&\bz_{3\times 1}\\s_a&\bz_{3\times
3}&k^\dag_a&\bz_{3\times 1}\\\bz_{1\times 3}&\bz_{1\times
3}&0&0\\k_a&\bz_{1\times
3}&0&0\eea\right)$ \\  & & \\
$D(3,1,1)$ & $\left (\bea{cccc} s_a & \bz_{3\times 3}& \bz_{3\times 3}&
\bz_{3\times 1}\\
\bz_{3\times 3}& s_a & \bz_{3\times 3}& \bz_{3\times 1}\\
 \bz_{3\times 3} & \bz_{3\times 3} & s_a& \bz_{3\times 1}\\
\bz_{1\times 3} & \bz_{1\times 3} & \bz_{1\times 3} & 0\eea\right)$ &
 $\left (\bea{cccc} \bz_{3\times 3} & \bz_{3\times 3}&\bz_{3\times 3} &
 \bz_{3\times 1} \\
s_a & \bz_{3\times
3} &\bz_{3\times 3} &\bz_{3\times 1} \\ \bz_{3\times 3} & s_a &
\bz_{3\times 3} & k^\dagger_a\\
-k_a & \bz_{1\times 3} & \bz_{1\times 3} & 0\eea\right)$\\
 & &  \\
 \hline
\end{tabular}
\caption{Vector representations: spin matrices
 $S_a$ and boost matrices $\eta_a$ where $s_a$ and $k_a$ are matrices (\ref{spin}) .}\label{vector}
 \end{table}

\medskip

Any set of matrices $S_a$, $\eta_a$ given in Table \ref{vector}
yields a finite-dimensional indecomposable representation of algebra $hg(1,3)$ which generates a 
representation of the extended Galilei
algebra via (\ref{galal}).
From our analysis, there exist ten indecomposable vector representations of $hg(1,3)$.

The finite-transformations corresponding to these realizations are found by integrating the Lie
equations for generators given in (\ref{galal}) and Table \ref{vector}. For a specific representation 
$D(n,m,\lambda)$ they have the following forms.

\medskip
\begin{itemize}

\item $D(0,0,0)$: The related representation space is formed by
 a field $\psi$ invariant under rotations and which transforms under a Galilean boost as
 \beq\label{0}\psi\to e^{\ri mf}\psi,
\eeq
where $f={\bf v\cdot x}+{\bf v}^2t/2$. Such transformations keep
invariant the bilinear form $I_1=\psi^*\psi$.

\smallskip

\item $D(1,0,0)$: The Galilean 3-vectors ${\bf R}=$column$(R_1, R_2, R_3)$
transforms as a vectors under rotations:
\[ {\bf R}\to{\bf R}\cos\theta+\frac{{\mbox{\boldmath$\theta$\unboldmath}}\times{\bf R}}{R}
\sin\theta+\frac{{\mbox{\boldmath$\theta$\unboldmath}}
({\mbox{\boldmath$\theta$\unboldmath}}\cdot{\bf R})}{\theta^2}(1-\cos\theta),\]
 and via
\beq\label{1}{\bf R}\to e^{\ri mf}{\bf R}, \eeq
under the Galilei boosts. Then, the bilinear form
$I_2={\bf R^*\cdot R}$ is invariant under Galilean transformations.

\smallskip

\item $D(1,1,0)$: Galilean 4-vector $\Psi_4=\left({\bf U},  \psi\right)$, where
${\bf U}$ is a vector and $ \psi$ is a scalar under the rotation
transformations. The action of Galilean boosts on $\bf U$ and $\psi$
 can be written as
\beq\label{2}\bea{l} {\bf U}\to e^{\ri mf}({\bf U}+{\bf v} \psi),\\
\psi\to e^{\ri mf}\psi.\eea\eeq

The corresponding invariant
of Galilei group is of the form $I_1=\psi^*\psi$.

\smallskip

\item $D(1,1,1)$: The second Galilean 4-vector
    $\tilde \Psi_4=\left({\bf R},B\right)$, where $\bf R$
 and $B$ transform under the Galilei boost as:
\beq\label{3}\bea{l}{\bf R}\to e^{\ri mf}{\bf R},\\ B\to e^{\ri mf}(B+
{\bf v\cdot R}).
\eea\eeq
The invariant form for these (and rotation) transformations can be written as $I_2={\bf R^*\cdot R}$.

\smallskip

\item $D(1,2,1)$: The Galilean 5-vector $\Psi_5=(\psi,{\bf U},C)$, where
$\psi$ and
 $C$ are scalars and ${\bf U}$ is a vector with respect to  rotations.
 Under a Galilean boost, they transform as:
 \beq\label{4}\bea{l} \tilde \psi\to e^{\ri mf}\tilde \psi,\\
{\bf U}\to e^{\ri mf}({\bf U}+{\bf v} \psi),\\
C\to e^{\ri mf}(C+{\bf v}\cdot{\bf U}+\frac 12{\bf v}^2\psi),\eea\eeq
and the invariants of these transformations are $I_2=\psi^*\psi$
 and $I_3=\psi^*C+\psi C^* -{\bf U}^*{\bf U}$ .

\smallskip

\item $D(2,0,0)$: The Galilean 6-vectors (bi-vectors) $\Psi_6=\left({\bf R,
W}\right)$, which under the Galilei boosts transform as:
 \beq\label{5}\begin{array}{l}
{\bf R}\to e^{\ri mf}{\bf R},\\
{\bf W}\to e^{\ri mf}({\bf W}+{\bf v\times R}),\end{array}\eeq
 and the corresponding invariants are $I_2={\bf R^*\cdot R}$ and $
 I_4={\bf R^*\cdot W}+{\bf
R\cdot W^*}$.

\smallskip

\item $D(2,1,0)$: The Galilean 7-vectors $\Psi_7=\left({\bf R,W},B\right)$. Under the Galilei boost its
components transform according to
\beq\label{31}\bea{l}{\bf R}\to e^{\ri mf}{\bf R},\\ B\to e^{\ri mf}(B+
{\bf v\cdot R}),\\
{\bf W}\to e^{\ri mf}({\bf W}+{\bf v\times R}).
\eea\eeq   The corresponding invariants are the same as the previous case:
 $I_2={\bf R}^*\cdot{\bf R}$ and $ I_4=
{\bf R^*\cdot W}+{\bf R\cdot W^*}$.
\smallskip

\item $D(2,1,1)$: The second 7-vector $\tilde \Psi_7=({\bf K,R},\psi)$
 with ${\bf K}$ being a 3-vector which under the Galilei boost transforms as
\beq\label{6}\bea{l}{\bf K}\to e^{\ri mf}\left({\bf K}+{\bf v\times R}+
{\bf v}\psi\right),
\eea\eeq and with $\bf R,\ \psi$ which transform as:
\beq\psi\to e^{\ri mf}\psi,\ {\bf R}\to e^{\ri mf}{\bf R}.\label{00}\eeq
The corresponding invariants are $I_1=\psi^*\psi$ and
$I_2={\bf R^*\cdot R}$.

\smallskip

\item $D(2,2,1)$: Galilean 8-vector $\Psi_8=\left({\bf R,K},B,\psi\right)$,
whose components with the Galilei boost transform as:
\beq\bea{l}\psi\to e^{\ri mf}\psi,\ {\bf R}\to e^{\ri mf}{\bf R},\\
{\bf K}\to e^{\ri mf}\left({\bf K}+{\bf v\times R}+{\bf v}\psi\right),\\
B\to e^{\ri mf}(B+{\bf v\cdot R})
.\eea\label{001}\eeq
 The
invariants of these transformations are $I_1=\psi^*\psi$,
$I_2={\bf R^*\cdot R}$ and $I_5=\psi^*B+\psi B^*-{\bf K\cdot R^*}-{\bf
K^*\cdot R}$.

\smallskip

\item $D(3,1,2)$: The ten-vector fields $\Psi_{10}=({\bf R, W, N},
B)$
combine three 3-vectors ${\bf R, W, N}$ and one scalar $B$. They
cotransform under the boost transformations via
\beq\label{8}\bea{l}{\bf R}\to e^{\ri mf}{\bf R},\\
{\bf W}\to e^{\ri mf}({\bf W}+{\bf v\times R}),\\
{\bf N}\to e^{\ri mf}\left({\bf N}+{\bf v\times W}+{\bf
v}B+{\bf v}({\bf v\cdot R})-\frac12{\bf v}^2\bf R\right),
\\B\to e^{\ri mf}(B+{\bf v\cdot R}).\eea\eeq
The
invariants of these transformations are $I_2={\bf R^*\cdot R}$, $ I_4=
{\bf R^*\cdot W}+{\bf R\cdot W^*}$ and $I_6=B^*B$ $+{\bf W\cdot W^*-
R\cdot
N^*}-{\bf N\cdot R^*}$. In Section 7, we shall utilize this vector
 in the Galilean electromagnetism.

\medskip

\end{itemize}
 Thus in addition to Galilean scalar $A$ there exist nine Galilean vectors
 enumerated in the above items.
 We see that the number of such vectors is notably larger then in the case of the proper Lorentz
 group, when there are only three irreducible multiplets whose components
 transform as vectors or scalars under rotations, namely,
 a four-vector  and  self-dual and anti self-dual
 components of an antisymmetric tensor \cite{corson}.

\section{Examples of Galilean vectors}

 In the previous section, we have described finite-dimensional indecomposable Galilean
vectors and present explicitly their group transformations and
invariants. They form the main tool for constructing
various Galilei invariant models. In particular, by using the realizations
 of matrices $S_a$ and $\eta_a$ given in Table \ref{vector}, it is
 possible to describe all non-equivalent Galilei
invariant first order partial differential equations for vector fields.
Here we present some important examples of Galilean vectors.

\medskip

\noindent{\em \underline Example 1.} Generators $P_0$ and $P_a$ of time and space
translations
defined by relations (\ref{galal}) and mass $m$ form a Galilean
5-vector of type $ \Psi_5$ provided we identify $P_0\sim C, {\bf P}\sim
{\bf U}, m\sim \psi$. For $m=0$ this 5-vector is reduced to a
4-vector of type $\tilde\Psi_4$.

\medskip

\noindent{\em \underline Example 2.}  Five-potential of Galilean electromagnetic
field
\cite{ijtp} $\hat A=(A_0, {\bf A}, A_4)$ with transformation law
\beq\bea{l} \bA'=\bA+\bv A_4,\\
A'_{0}=A_0+\bv\cdot\bA +\frac 12\bv^2 A_4,\\ A'_{4}=A_4
\eea\label{atransf}\eeq is an example of Galilean 5-vector
field with zero mass, which is a carrier space of the representation $D_5$
described in Table \ref{vector}.

\medskip

\noindent{\em \underline Example 3.}  The field strength tensor of Galilean
electromagnetic field
\beq F_{\mu\nu} =
\partial_\mu A_\nu - \partial_\nu A_\mu \label{n1}\eeq
where $\mu, \nu=0,1,2,3,4$ and by definition $\partial_4 A_\mu=0$,
is the example of massless 10-vector which transforms in accordance with
(\ref{8}).
The explicit
relation between components of $F_{\mu\nu}$ and $\Psi_{10}$ is
given by the following formula:
\beq F_{\mu\nu}=\left (\bea{ccccc}
0 & -N_1 & -N_2 & -N_3 & B\\
N_1 & 0 & W_3 & -W_2 & R_1\\
N_2 & -W_3 & 0 & W_1 & R_2\\
N_3 & W_2 & -W_1 & 0 & R_3\\
-B & -R_1 & -R_2 & -R_3 & 0\eea \right
)\label{fmunu}\eeq
 where
\beq \label{F}\bea{l}
B=\pd_0 A_4,\\
{\bf W}=\bfn\times\bA,\\
{\bf N}=\bfn A_0-\pd_0\bA,\\
 {\bf R}=\bfn A_4.\eea \eeq

 Some subsets of components of $F_{\mu\nu}$ form carrier spaces  for other
 representations of the Galilei group. For example, if $R_a=F_{4a}\equiv 0$
 then the
 remaining components of $F_{\mu\nu}$ transform as a seven-vector of type
 $\Psi_7$.
 The complete list of various vectors which can be formed using
 components of $F_{\mu\nu}$ is present in Section 5.

\medskip

\noindent{\em \underline Example 4.} As another example of a Galilean field can be considered the 
matrix 5-vector $D_5=(\psi,U_1,U_2,U_3, C)$ with
components
\beq\label{dirac}\bea{l}
\psi=\beta_0=\sqrt2(\gamma_0+\gamma_4),\\
 C=\beta_4=\sqrt2 (\gamma_0-\gamma_4),\\
  {U_a}=\beta_a=\gamma_a
\eea\eeq
where $\gamma_\mu$ are Dirac matrices,
$\mu=0,\dots,4,\ a=1,2,3$. Such set of matrices commutes as a 5-vector with
the Galilei generators (\ref{cov}) if we choose
\beq\label{mat}\bea{l}
S_a=\frac14\varepsilon_{abc}\gamma_b\gamma_c,\\
\eta_a=\frac{1}{2\sqrt2}\left(\gamma_0+\gamma_4\right)\gamma_a.\eea
\eeq

\medskip

\noindent{\em \underline Example 5.} As in the two previous examples, the matrix
tensor
\beq\label{tensor}
S_{\mu\nu}=\beta_\mu\beta_\nu-\beta_\nu\beta_\mu\eeq transforms
like $F_{\mu\nu}$, that is, as a 10-vector. Moreover, it is possible
to form the following matrix vectors:
    \beq\bea{l}\tilde \Psi_4: \ \left( R_a= S_{4a},\ B=S_{40}\right),\\
\tilde \Psi_6:\ \left(V_a=\frac12\varepsilon_{abc}S_{bc}, \ W_a=S_{4a}\right), \\
\tilde \Psi_7:\ \left(K_a= S_{0a},\ R_a= \frac12\varepsilon_{abc}S_{bc},\
A=S_{40}\right).\eea\label{psi}\eeq

\section{Contractions of representations of Lorentz algebra}

It is well known that the Galilei group (algebra) and (some of) its representations  can be obtained 
from the Poincar\'e group (algebra) and from its appropriate representations by a limiting procedure 
called "contraction". The process of contraction has, by now, an extensive literature. First it was 
proposed by Segal \cite{segal} and in more specific forms by In\"on\"u and Wigner \cite{contraction}, 
by Saletan \cite{saletan}, by Doebner and Melsheimer \cite{doebner} and many others,
see the excellent review article of L\^ohmus \cite{loh} and references cited therein, a more recent 
review can be found in \cite{weimar}.

The Lie algebra of a given Lie group is defined via commutation relations for basis elements. 
As shown by Cartan, whenever we change the basis by a non-singular transformation we come to algebra 
isomorphic to the original one. However, if the transformation is singular, a new algebra may be received,
 provided this transformation leads to well defined commutation relations for the transformed basis 
 elements.

In the simplest case
a {\it contraction} is a limit procedure which transforms an
 $N$-dimensional Lie algebra ${\cal L}$ into an non-isomorphic Lie
 algebra ${\cal L}'$, also with $N$ dimensions.
 The commutation relations of a {\em contracted Lie algebra}
 ${\cal{L}}'$ are given by:
\begin{equation}
[x,y]'\equiv\lim_{\varepsilon\rightarrow\varepsilon_0}
 {W}_\varepsilon^{-1}([{W}_\varepsilon(x),
 {W}_\varepsilon(y)],
\label{contractiondefinition}\end{equation}
where ${W}_\varepsilon\in\ $GL($N,k$) is a non-singular linear transformation
 of ${\cal{L}}$, with $\varepsilon_0$ being a
 singularity point of its inverse ${W}_\varepsilon^{-1}$.

The papers \cite{segal}-\cite{weimar} indicate different ways of performing this (and more general) 
singular transformation and necessary and sufficient conditions that a given Lie algebra can be 
contracted into another one. However there is no a regular way to obtain representations of the 
contracted algebra starting with the representations of the initial one. Namely, in contracting 
representations we meet the following difficulties of principle:

i) Contracting the faithful representation of a given Lie algebra we obtain in general a non-faithful 
representation of the resulting Lie algebra since a part of generators is represented trivially.

ii) The resulting (contracted) algebra is always non-compact. Hence any contraction of a Hermitian 
irreducible representation of some compact Lie algebra, which is always {\it finite}-dimensional, 
has to yield at the end to an {\it infinite}-dimensional Hermitian irreducible representation of 
the non-compact Lie algebra.

In\"on\"u and Wigner in \cite{contraction} mentioned possible ways of treating the difficulties 
(for details see paper \cite{micke} where contraction of representations of the de Sitter group 
was carried out), one of them will be used in what follows.

There is
a simple contractions procedure (the In\"on\"u-Wigner contraction) connecting the Lie algebra $so(1,3)$ 
of Lorentz group  with algebra $hg(1,3)$.
The related transformation $W$ does not change basis elements of $so(1,3)$ forming its subalgebra $so(3)$  
while the remaining basis elements are multiplied by a small parameter $\varepsilon$ which tends to zero 
\cite{contraction}.

Here we find representations of the Lorentz group whose contraction makes it possible to obtain found 
realizations of homogeneous Galilei group. A specific feature of these representations is that they are 
in general case completely reducible while the corresponding contracted representations of Galilei algebra 
are indecomposable ones.

Let $S_{\mu\nu}, \ \mu, \nu =0,1,2,3$ are matrices realizing a representation of the Lorentz algebra, 
i.e., satisfying the relations
\beq\label{13}[S_{\mu\nu}, S_{\lambda\sigma}]=i(g_{\mu\lambda}S_{\nu\sigma}+g_{\nu\sigma}
S_{\mu\lambda}-g_{\nu\lambda}S_{\mu\sigma}-g_{\mu\sigma}S_{\nu\lambda})\eeq
with $g_{\mu\nu}=\texttt{diag}(1,-1,-1,-1)$. The contraction procedure consists in transition to a new 
basis
\[S_{ab}\to S_{ab}, \ S_{0a}\to\varepsilon S_{0a}\]
and simultaneous transformation of all basis elements  $S_{\mu\nu}\to S'_{\mu\nu}=US_{\mu\nu}U^{-1}$
with a matrix $U$ depending on $\varepsilon$. Moreover, $u$ should depend on $\varepsilon$ in a tricky way, 
such that all the transformed generators $S'_{\mu\nu}$ are kept non-trivial
when $\varepsilon\to 0$  \cite{contraction}.

We suppose that representations obtained by the contraction are indecomposable representations 
$D(n,m,\lambda)$ described in Section 4.
To construct representations of Lorentz algebra which can be contracted to representations 
$D(n,m,\lambda)$ we use the following observation.

{\bf Lemma}  {\it Let $\{S_a,\eta_a\}$ be an indecomposable set of
matrices realizing one of representations $D(1,1,0), D(1,1,1), D(1,2,1), D(2,0,0)$ or $D(3,1,1)$ 
presented in Table 1. Then matrices
\beq\label{12}S_{ab}=\varepsilon_{abs}S_c,\ S_{0a}=\nu(\eta_a-
\eta^\dag_a)\eeq
where $\nu=1$ for representations $D(1,1,0), D(1,1,1)\ D(2,0,0)$ and
$\nu=\frac1{\sqrt2}$ for representations $D(1,2,1), \ D(3,1,1)$,
form a basis of the Lie algebra of Lorentz group.}

The proof is reduced to the direct verification, that for all basis
elements $S_a,\ \eta_a $  of homogeneous Galilei algebra the
corresponding linear combinations (\ref{12}) satisfy the relations
(\ref{13}), i.e., do form a basis of the Lorentz algebra.

For representations $D(1,0,0)$, $(D(2,1,1), D(2,1,0))$ and $D(2,2,1)$ the related matrices
(\ref{12}) do not form
a basis of Lie algebra. Nevertheless it is possible to find the
corresponding generators of the Lorentz algebra starting with its known
representations of dimension $3\times3, \ 7\times7$ and $8\times 8$
correspondingly. We choose these representations in the following forms:
\beq\label{141}S_{ab}=\varepsilon_{abc}s_c,\ S_{0a}=is_a,\eeq
\beq\label{15}S_{ab}=\varepsilon_{abc}\left(\bea{ccc}s_c&\bz_{3\times3}&
\bz_{3\times1}\\\bz_{3\times3}&s_c&\bz_{3\times1}\\
\bz_{1\times3}&\bz_{1\times3}&0\eea\right),\
S_{0a}=\frac12\left(\bea{ccc}is_a&-s_a&i\sqrt2k^\dag_a\\s_a&is_a&-
\sqrt2k^\dag_a\\i\sqrt2k_a&\sqrt2k_a&0\eea\right),\eeq
and
\beq\label{16}S_{ab}=\varepsilon_{abc} \left(\bea{cccc}s_c&\bz_{3\times 3}&\bz_{3\times
1}&\bz_{3\times 1}\\\bz_{3\times 3}&s_c&\bz_{3\times
1}&\bz_{3\times 1}\\\bz_{1\times 3}&\bz_{1\times 3}&0&0\\
\bz_{1\times 3}&\bz_{1\times 3}&0&0\eea\right),\ S_{0a}=\frac12\left(\bea{cccc}is_a&-s_a&i
k^\dag_a&k^\dag_a\\s_a&is_a&-k^\dag_a &ik^\dag_a\\ik_a&k_a&0&0\\-k_a&ik_a&0&0\eea\right)\eeq
where $s_a$ and $k_a$ are matrices given by equations (\ref{spin}) and (\ref{k}).

Matrices (\ref{12})-(\ref{16}) form basises of representations of Lorentz algebra which are in general 
reducible.  These representations (together with the related realizations of
homogeneous Galilei algebra given in Table 1) are enumerated in Table 2.
\begin{table}
\begin{tabular}{|c|c|c|c|}\hline
Representation&Representations&Basis & Contracting matrix $U$ \\of algebra $hg(1,3)$&of 
algebra $so(1,3)$&elements &\\&&$S_{\mu\nu}$&\\
 \hline
$D(1,0,0)$&$D(1,0)$&(\ref{141})&$I_{3\times3}$\\&&&\\
$D(1,1,0)$&$D(\frac12, \frac12)$&(\ref{12})&$\left(\bea{cc}I_{3\times3}&
\bz_{3\times1}\\\bz_{1\times3}&\varepsilon^{-1}\eea\right)$\\&&&\\
$D(1,1,1)$&$D(\frac12, \frac12)$&(\ref{12})&$\left(\bea{cc}I_{3\times3}&
\bz_{3\times1}\\\bz_{1\times3}&\varepsilon\eea\right)$\\&&&\\
$D(1,2,1)$&$
 D(\frac12, \frac12)\oplus D(0, 0)$&(\ref{12})&$\left(\bea{ccc}I_{3\times3}
 &\bz_{3\times1}&\bz_{3\times1}\\\bz_{1\times3}&\varepsilon&0\\
 \bz_{1\times3}&0&\varepsilon^{-1}\eea\right)$\\&&&\\
 $D(2,0,0)$&$
  D(0,1)\oplus D(1,0)$&(\ref{12})&$\left(\bea{cc}I_{3\times3}&
  \bz_{3\times3}\\\bz_{3\times3}&\varepsilon^{-1} I_{3\times3}\eea\right)$
   \\&&&\\
   $D(2,1,0) $&$
 D(\frac12, \frac12)\oplus D(1,0)$&(\ref{15})&$\left(\bea{ccc}
 I_{3\times3}&\bz_{3\times3}&\bz_{3\times1}\\\bz_{3\times3}&\varepsilon^{-1}
  I_{3\times3}&\bz_{3\times1}\\\bz_{1\times3}&\bz_{1\times3}&\varepsilon^{-1}\eea\right)$\\&&&\\$D(2,1,1) 
  $&$
 D(\frac12, \frac12)\oplus D(1,0)$&(\ref{15})&$\left(\bea{ccc}
 I_{3\times3}&\bz_{3\times3}&\bz_{3\times1}\\
 \bz_{3\times3}&\varepsilon I_{3\times3}&\bz_{3\times1}\\
 \bz_{1\times3}&\bz_{1\times3}&\varepsilon \eea\right)$\\&&&\\
 $D(2,2,1)$&$\bea{l}
  D(\frac12, \frac12)\oplus\\ D(0,0)\oplus D(0,0)\eea$&(\ref{16})&$\left(\bea{cccc}I_{3\times3}&
  \bz_{3\times3}&\bz_{3\times1}&\bz_{3\times1}\\\bz_{3\times3}
  &\varepsilon^{-1}I_{3\times3}&\bz_{1\times3}&\bz_{1\times3}\\
  \bz_{1\times3}&\bz_{1\times3}&1&0\\\bz_{1\times3}&\bz_{1\times3}&0&
  \varepsilon^{-1}\eea\right)$\\&&&\\
  $D(3,1,1)$&$\bea{ll} D(0,1)\oplus D(1,0)
 \\ \oplus D(\frac12, \frac12)\eea$&(\ref{12})&$\left(\bea{cccc}I_{3\times3}
 &\bz_{3\times3}&\bz_{3\times3}&\bz_{3\times1}\\\bz_{3\times3}&
 \varepsilon^{-1}I_{3\times3}&\bz_{3\times3}&\bz_{3\times1}\\\bz_{3\times3}&\bz_{3\times3}&
 \varepsilon^{-2}I_{3\times3}&\bz_{3\times1}\\\bz_{1\times3}&\bz_{1\times3}&
 \bz_{1\times3}&\varepsilon^{-1}\eea\right)$\\\hline\end{tabular}
 \caption{Representations of Lorentz algebra and contracting matrices}\label{contraction}\end{table}

 Thus we obtain the indecomposable representations of the Lie algebra of the homogeneous Galilei group 
 starting with finite-dimension representations of the Lorentz algebra $so(1,3)$ and applying the 
 contraction procedure. Special features of this approach are summarized at the following items.

\medskip

\begin{itemize}
\item To obtain {\it indecomposable} representations of algebra $hg(1,3)$ found in Section 4 we were 
supposed to use {\it completely reducible} representations of the Lorentz algebra.
\item It is possible to obtain different non-equivalent realizations of $hg(1,3)$ starting with a given 
representation of $so(1,3)$. For example, both realizations $D_7$ and $D_8$ can be obtained via 
contractions of the representation $D(\frac12,\frac12)\oplus D(1,0)$.
\item It is possible to obtain a given representation of $hg(1,3)$ via contraction of different 
representations of $so(1,3)$. For example, above mentioned representation 
$D(\frac12,\frac12)\oplus D(1,0)$ can be replaced by $D(\frac12,\frac12)\oplus D(0,1)$, or, 
more generally, any representation $D(m,n)$ of $so(1,3)$ can be replaced by $D(n,m)$.

\end{itemize}

\section{Galilean linear spin one-half
wave equation with Pauli anomalous interaction} 
The completed list of indecomposable spinor and vector representations of algebra $hg(1,3)$ found in 
Section 3 and 4 can be used to construct Galilei invariant models for spinorial and vector fields both 
linear and nonlinear ones. In particular these representations can be applied to derive systems of 
equations invariant with respect to Galilei group.

In this section we apply them to describe possible Pauli type interactions for spinor field invariant 
with respect to Galilei group.

\subsection{Reduction approach}

Consider the Dirac equation describing a fermionic field $\Psi(x)$
coupled to a gauge field $A_\mu$, and a Pauli anomalous term :
\beq \label{Di}\left (\gamma_\mu\pi^\mu+k[\gamma^\mu,\gamma^\nu]
F_{\mu\nu}-\lambda \right )\Psi(x)=0, \label{diracequation}\eeq
where \[ \pi_\mu\equiv p_\mu-qA_\mu, \] and the second term is
the Pauli anomalous term.

A natural way to construct a Galilean analogue of equation
(\ref{diracequation}) is to generalize it to the case of $(4,
1)$-dimensional space and then make the reduction discussed in
Section 1. However in this case it is desirable to present a clear
physical interpretation for all values obtained by the reduction.
In addition, in this way we cannot obtain the most general Pauli
interaction term. Thus we consider two possibilities: the
reduction of equation defined in $(4,1)$-dimensional space and a
direct search for Pauli term invariant under the Galilei group.

In order to perform the reduction,
  it is sufficient to change in (\ref{Di}) the
Dirac matrices to their Galilean analogues, i.e., to change $\gamma_\mu\to
\beta_\mu$ where
$\beta_\mu$ are defined by relations (\ref{dirac}). In particular they
may be chosen
as \cite{ourletter}
\beq
\beta^{0}=\left(
\begin{array}{cc}
0&0\nonumber\\
-\sqrt{2}&0
\end{array}\right),\qquad
\beta^a=
\left(\begin{array}{cc}
\sigma^a&0\\
0&-\sigma^a
\end{array}\right),\qquad
\beta^4=\left(
\begin{array}{cc}
0&\sqrt{2}\\
0&0
\end{array}\right). \label{gamma}
\eeq
If we substitute these matrices into equation
(\ref{diracequation}) for $\psi=\left(\bea{c}\phi\\\chi\eea\right)$
(where $\phi$
and $\chi$ are two component spinors) and use the notations (\ref{fmunu}),
 we find
 \beq\bea{l}
({\mbox{\boldmath$\sigma$\unboldmath}}\cdot{\mbox{\boldmath$\pi$\unboldmath}}
-\lambda+4\ri k{\mbox{\boldmath$\sigma$\unboldmath}}\cdot\bW+4kB]\varphi+
\sqrt{2}\ (\pi_4+4k{\mbox{\boldmath$\theta$\unboldmath}}\cdot\bR)\chi=0,\\
{\sqrt{2}\ (\pi_0-4k{\mbox{\boldmath$\sigma$\unboldmath}}\cdot\bN)}
\varphi+
({\mbox{\boldmath$\sigma$\unboldmath}}\cdot{\mbox{\boldmath$\pi$\unboldmath}}
+\lambda-4\ri k{\mbox{\boldmath$\sigma$\unboldmath}}\cdot\bW+4kB)\chi=0\eea
\label{diracequationscomponents}\eeq
where $\pi_4=m-qA_4$. Solving the first of equations
(\ref{diracequationscomponents})
for $\chi$ and substituting the solution into the second line we come to
the Schr{\"o}dinger equation with spin dependent potential which is not discussed here.

In the magnetic limit, defined as in reference \cite{ijtp}:
 \beq\label{ml}
(\bA_m, \phi_m)\ho A_m= \left (A_0, \bA_m, A_4\right )=\left
(-\phi_m, \bA_m, 0\right ),
\eeq
we find
\[\bea{l}
\bB_m={\bf W}=\bfn\ti\bA_m,\\
 \bE_m={\bf N}=-\bfn\phi_m-\pd_t\bA_m.\eea
\]
 Note that ${\bf R}=\bz$ and $B=0$, and
Galilei transformations for $\bf A_m$, $\phi_m$, $\bB_m$ and $\bE_m$
have the form
\beq\begin{array}{ll}
\bA_m\to\bA_m,\qquad & \phi_m\to\phi_m-{\bf v}\cdot\bA_m,\\
 \bB_m\to \bB_m,\qquad & \bE_m\to\bE_m+{\bf v}\times\bB_m.\label{EH}
\end{array}\eeq
With these definitions, equation (\ref{diracequationscomponents})
becomes
\[\bea{l}
({\mbox{\boldmath$\sigma$\unboldmath}}\cdot({\mbox{\boldmath$\pi$\unboldmath}}
-q\bA_m)-\lambda+4\ri k{\mbox{\boldmath$\sigma$\unboldmath}}\cdot\bB_m)
\varphi+
\sqrt{2}\ m \chi=0,\\
{\sqrt{2}\ (p_0+q\phi_m-4k{\mbox{\boldmath$\sigma$\unboldmath}}
\cdot\bE_m)}\varphi+
[{\mbox{\boldmath$\sigma$\unboldmath}}\cdot({\mbox{\boldmath$\pi$\unboldmath}}
-q\bA_m)+\lambda-4\ri k{\mbox{\boldmath$\sigma$\unboldmath}}\cdot\bB_m]\chi
=0.\eea
\]
This may be rewritten as
\beq\label{elim}
\left[\gamma_\mu\pi^\mu-\lambda+2k\left({\bf
S}\cdot \bB_m+{\mbox{\boldmath$\eta$\unboldmath}}\cdot\bE_m\right)\right]\psi=0
\eeq where
$\bf S$ and ${\mbox{\boldmath$\eta$\unboldmath}}$ are matrix vectors whose components are
given in equation (\ref{mat}).

Thus we obtain in rather elegant way equation (\ref{elim}) which
is manifestly Galilei invariant and describes both the minimal and
anomalous couplings of particle of spin 1/2 with an external
Galilean electromagnetic field of magnetic type. In the case $k=0$, i.e.,
when only the minimal interaction is present, this equation is reduced
to the L\'evy-Leblond equation. However, it
presents only one of many possibilities to introduce anomalous
coupling into the L\'evy-Leblond equation. A more general approach
is presented in the following subsection.

\subsection{Direct approach}

Anomalous Pauli interaction is represented in equation (\ref{Di})
by the term $k[\gamma_\mu, \gamma_\nu] F^{\mu\nu}$, which is linear
 with respect to the electromagnetic field strength $F$. It is possible to show
that the requirement of relativistic invariance defines this term
in a unique way (up to a value of the coupling constant $k$).

In contrast with the above, in the Galilei-invariant approach
there are more possibilities to introduce the anomalous
interaction. In this subsection we use our knowledge of Galilean
vector field (refer to Section 4) to find the most general form of
anomalous interaction for Galilean spinors.

First we remind that there exist two types of massless Galilean
 4-vector fields, i.e., $\Psi_4$ and $\tilde \Psi_4$, whose transformation
properties are defined by equations (\ref{2}) and (\ref{3}) with $m=0$, and a
5-vector $\Psi_5$ which transforms as given in (\ref{atransf}).
In other words we have three types of potentials of external
vector field which can be used to introduce the {\it minimal}
interaction into the L\'evy-Leblond equation. And they are
potentials which generate field strengthes involved into anomalous
interaction terms whose examples are present in the previous subsection.

We notice that in addition it is possible to introduce minimal and
anomalous
interactions with fields whose potentials are three-vectors or scalars.
Moreover,
the Galilei invariance condition admits some constrains for the potentials which
generate additional possible anomalous interactions.

Let us start with the vector $\tilde\Psi_4$ which corresponds to the
magnetic limit field considered in the previous section, and find
the most general Galilean scalar matrix $F$ linear in $\bB_m$ and
$\bE_m$. Expanding $F$ via the complete set of matrices
$\beta_\mu$ (\ref{dirac}) and $ S_{\mu\nu}$ (\ref{tensor}) with
well defined transformation properties and using (\ref{EH}) we
easily find that $F=2k\left({\bf S}\cdot
\bB_m+{\mbox{\boldmath$\eta$\unboldmath}}\cdot\bE_m\right)+
g{\mbox{\boldmath$\eta$\unboldmath}}\cdot\bB_m$ where $k$
and $g$ are arbitrary parameters. In other words, we come to the
following Galilei-invariant equation for spinor field with the
minimal and anomalous interaction:
\beq\label{pauliterm}
\left [\gamma_\mu\pi^\mu-\lambda+2k\left({\bf S}\cdot
\bB_m+{\mbox{\boldmath$\eta$\unboldmath}}\cdot\bE_m\right)+
g{\mbox{\boldmath$\eta$\unboldmath}}\cdot\bB_m\right]\psi=0.\eeq
In contrast with (\ref{elim}), equation (\ref{pauliterm}) includes
two coupling constants, $k$ and $g$.

The other Galilean limit, named `electric limit' \cite{lebellac}
for the electromagnetic field corresponds to the gauge fields of
type $\Psi_4$, i.e., the related vector-potential
 \beq\label{el}
(\bA_e, \phi_e)\ho A_e= \left (A_0, \bA, A_4\right )=\left
(0, \bA_e, \phi_e\right ),
\eeq
 and field strengthes $\bB_e=\bfn\times\bA_e,
\ \bE_e=-\bfn\phi_e, \Phi=\partial_0\phi$ transform as
\beq\label{Ae}\begin{array}{l}
\bA_e\to \bA_e+{\bf v}\phi_e,\qquad  \phi_e\to\phi_e\\
 \bE_e\to\bE_e,\qquad \bB_e\to \bB_e-{\bf v}\times\bE_e,\ \ \Phi\to \Phi-
 {\bf v}\cdot{\bf E_e}.
\end{array}\eeq

Searching for the related Galilean Dirac equation with general
Pauli interaction term and using the fact that the vectors
$(\bB_e, -\bE_e)$ have the same transformation properties as
$(\bE_m, \bB_m)$ we conclude that to achieve our goal it is
sufficient to change $\bE_m\to\bB_e $ and $\bB_m\to-\bE_e$ in
(\ref{pauliterm}) and add the additional invariant term $\gamma_0\Phi +
{\mbox{\boldmath$\gamma$\unboldmath}}\cdot{\bf E}_e$. In addition, to keep the Galilei invariance
we should introduce the minimal interaction in the following manner:
\[\bea{l}
\pi_0=\ri\partial_0,\\
 \pi_a=-\ri\partial_a-q(A_e)_a,\\
\pi_4=m+q\phi_e
\eea\]
As a result we obtain
\beq\label{pauliterm1}\left(\gamma_\mu\pi^\mu-
\lambda+2k\left({\mbox{\boldmath$\eta$\unboldmath}}
\cdot\bB_e
-{\bf S}\cdot \bE_e\right)+g{\mbox{\boldmath$\eta$\unboldmath}}
\cdot\bE_e+r(\gamma_0\phi_e +
{\mbox{\boldmath$\gamma$\unboldmath}}\cdot{\bf E}_e)\right)\psi=0.\eeq

Thus there exist at least two ways to describe anomalous interaction in the Galilei invariant approach,
 presented by equations (\ref{pauliterm}) and (\ref{pauliterm1}). Equation (\ref{pauliterm}) includes 
 two coupling constants while in (\ref{pauliterm1}) the number of such constants is equal to 3.

Note that transformation laws (\ref{EH}) enable to impose in (\ref{pauliterm}) the
Galilei-invariant condition ${\bf A}_m=0$ on the vector-potential.
Thus there exist one more Galilei-invariant equation with
anomalous interaction, namely
\beq\label{paulismall}\left[\gamma_0(p_0+q\phi)-
{\mbox{\boldmath$\gamma$\unboldmath}}\cdot{\bf p}-\gamma_4m+
2k{\mbox{\boldmath$\eta$\unboldmath}}\cdot\bfn \phi\right]\psi=0.\eeq

Analogously, starting with (\ref{pauliterm1}) taking into account that transformations (\ref{Ae})
are compatible with the condition $\phi_e=0$ we come to one more
Galilei invariant equation with Pauli interaction term, namely
\beq\label{paulismall1}\left(\gamma_0\partial_0-
{\mbox{\boldmath$\gamma$\unboldmath}}\cdot
{\mbox{\boldmath$\pi$\unboldmath}}-\gamma_4m-
\lambda
+g{\mbox{\boldmath$\eta$\unboldmath}}\cdot\bB\right)\psi=0,\eeq where
${\mbox{\boldmath$\pi$\unboldmath}}=-
\ri\nabla-q{\bf A},\ \bB=\bfn\times \bf A$.

Gauge invariant fields $\bE=-\bfn\phi$ and $\bB=\bfn\times \bf A$
present in equations (\ref{paulismall}) and (\ref{paulismall1}) can
be interpreted as external electric and magnetic fields.

Consider also the case when the potential of external field form a
5-vector, and related field strengthes are given by formulae
(\ref{n1})-(\ref{F}). We will not discuss possible physical
interpretation of the corresponding ten-component vector field but
mention that there exist a formal possibility to introduce the
anomalous interaction with it by generalizing the Galilean Dirac
equation to the following one:
\beq \label{Dian}\left
(\gamma^\mu\pi_\mu+k[\gamma^\mu,\gamma^\nu]
F_{\mu\nu}+g\eta_aR_a+r(S_aR_a+\eta_aU_a)-\lambda \right)\Psi(x)=0,
\eeq with three coupling constants $k,\ g\ r$ additional
to $q$.

Like equation (\ref{diracequation}) with $\gamma$-matrices
(\ref{gamma}), equation (\ref{Dian}) is Galilei-invariant but is
quite more general.

Finally let the external field is defined by a five-vector potential
satisfying the
Galilei invariant constraint
\beq\label{vec}\partial_aA_4=0,\eeq
then $F_{4a}\equiv 0$ and tensor $F_{\mu\nu}$ (\ref{fmunu}) is reduced to
the
seven-vector $\tilde\Psi_7=({\bf K},{\bf R},A)$ whose components are
\[{\bf K}={\bfn}A_0-\partial_0{\bf A},\ {\bf R}=\bfn\times{\bf A},\
A=\partial_0A_4.\]

To find the corresponding invariants linear in $F_{\mu\nu}$ it is
necessary to construct
invariant scalar products of vector functions (\ref{vec}) and matrices
(\ref{psi})
belonging to $\tilde\Psi_7$. As a result we come to the following equation
\beq \label{Dian1}\left
(\gamma^\mu\pi_\mu
+g\eta_aR_a+r(S_aR_a+\eta_aK_a+S^{04}A)+\nu\gamma_0A-\lambda \right)
\psi(x)=0,
\eeq
where $g,r$ and $\nu$ are coupling constants. For $A=0$ equation
(\ref{Dian1}) is reduced
to equation (\ref{pauliterm}) which describe anomalous interaction of the
Galilei particle
of spin 1/2 with external field of magnetic type.

Formulae (\ref{pauliterm}), (\ref{pauliterm1}), (\ref{paulismall}),
(\ref{paulismall1}),
(\ref{Dian}) and (\ref{Dian1}) present all non-equivalent Galilean
invariant
equations for spinor field,
describing minimal and anomalous interactions with external gauge fields.

We see that the direct search for Galilei invariant Pauli
interaction makes it possible to find more general coupling than
the reduction method.

\subsection{Galilean system with spin-orbit coupling}

 In this section we consider one of the described systems and discuss its physical content.

Let us start with equation (\ref{paulismall}) which describes interaction
of the Galilean
spinor particle with an external electric field. Choosing
$\gamma$-matrices in the form
(\ref{gamma}) and denoting $k=-\frac{q\hat k}{4m}$ we write this equation componentwise
\[\bea{l}
({\mbox{\boldmath$\sigma$\unboldmath}}\cdot\bp-\lambda)\varphi+
\sqrt{2}\ m \chi=0,\\
\sqrt{2}\ (p_0+q\phi+\frac{q\hat k}{m}{\mbox{\boldmath$\sigma$
\unboldmath}}\cdot{\bf E} )\varphi+
({\mbox{\boldmath$\sigma$\unboldmath}}\cdot{\bp}+\lambda)\chi=0\eea
\]
where ${\bf E}=-\nabla\phi$.

Solving the first equation for $\chi$ and substituting the
result
into the second equation we obtain
\beq \label{m6}L\varphi\equiv\left( p_0- \frac{p^2+\lambda^2}{2m}+q\phi-
\frac{q\hat k}{m}{\mbox{\boldmath$\sigma$\unboldmath}}\cdot{\bf E} \right)\varphi=0.\eeq

In other word we come to the Galilei invariant Schr\"odinger equation with a matrix
potential.

To analyze the physical content of equation (\ref{m6}) we transform it to
a  more familiar
 form using the operator $U=\exp(-\frac{ki}{m}
 {\mbox{\boldmath$\sigma$\unboldmath}}\cdot{\bf p})$. Applying this operator
 to $\varphi$ and $L$ we obtain the equation $L'\varphi'=0$ where
 $\varphi'=U\varphi$,
\[L'=ULU^{-1}=p_0-\frac{p^2}{2m}+q\phi-\frac{\lambda^2}{2m}-
\frac{q\hat k^2}
{2m^2}
\left({\mbox{\boldmath$\sigma$\unboldmath}}\cdot
({\bf p}\times{\bf E}-{\bf E}\times{\bf p})-\texttt{div} {\bf E}\right)+\cdots\]
where the dots denote the terms of order $o(\frac1{m^3})$.

All terms in operator $L'$ have exact physical meaning. In particular,
the last two terms describe
spin-orbit and Darwin coupling of a Galilean particle with an external
field.

 \section{Discussion}

 The relativity principle is one of the corner stones of modern physics.
 Moreover, it should be applied while considering phenomena characterizing
 by circumlight velocities but also in the cases when the velocities are
 small in comparison with the velocity of light. In the latest case any
 well formulated physical theory should not be simple "non-relativistic",
 but to satisfy the Galilei relativity principle. In other words, the
 group of motion of the special relativity (i.e., the Lorentz group) should
 be replaced by the Galilei group.

 It appears that a consistent use of Galilei group and its representations
 in many aspects is much more complicated than in the case
 of Poincar\'e group. In particular, description of finite-
 dimensional representations of the homogeneous Galilei group is a wild
 algebraic problem while such representations for the Lorentz group have
 been found long time ago.

 The main goal of the present paper was to present the completed
 classification of indecomposable Galilean fields which
 transform as vectors or are scalars with respect to the rotation
 transformations. In other words, we present the classification of
 indecomposable finite dimension
 representations of the homogeneous Galilei group which being reduced
 to the rotation subgroup correspond to spins $s\leq1$. In contrast with
 the fields with
 spin $s>1$ the Galilean vectors field can be described completely. The
 results of this
 classification are presented in Section 3.

 In contrast with the Lorentz vectors, the number of non-equivalent
 Galilean vectors appears
 to be rather extended. Namely, for the Lorentz group there exist the
 following indecomposable vector fields: four-vector, bi-vector (i.e.,
 antisymmetric tensor of second rank) and two three-vectors which are
 nothing but a self-dual and anti self-dual parts of the antisymmetric
 tensor. In the case of Galilei group it is possible to indicate nine
 indecomposable vector fields.

  We use our knowledge of vector representations of the Galilei group to
 describe all possible
 Pauli interactions for spinor fields, compatible with the
 Galilei invariance. The number
 of such interactions appears to be rather extended in contrast
 with the relativistic
 approach where this interaction is unique up to the coupling constant.
 We show that there exist such Galilei invariant systems which describe spin-orbit and Darwin couplings 
 which are traditionally treated as pure relativistic effects.

 It is generally accepted to think that Galilei group and its representations
can be obtained easily
starting with representations of the Lorentz group and making the
In\"on\"u-Wigner contraction \cite{contraction}. We had shown that this
procedure is not too straightforward in as much as starting with
indecomposable representations of the Lorentz group we can obtain only
a part of the corresponding representations of homogeneous Galilei group. On the other hand,
it is possible to contract reducible representations of the Lorentz group to indecomposable 
representations of the homogeneous Galilei group.

 The completed list of vector-scalar and spinor indecomposable representations representations
  presented in Sections 3 and 4 opens a way to describe all non-equivalent systems of Galilei 
  invariant equations for scalar, spinor and vector fields.
 We plane to discuss these equations in detail in the following paper.

\section*{Ackowledgements} This publication is based on work sponsored by the Grant Agency of
the Academy of Sciences of the Czech Republic under project Number
A1010711. Partial financial support was provided
 by the Natural Sciences and Engineering Research Council (NSERC)
 of Canada.

\end{document}